\documentclass[fleqn,12pt,twoside]{article}
\usepackage{espcrc1,amsmath,amssymb,graphicx,multirow,subfigure,oldgerm,euscript,mathrsfs}

\usepackage{graphicx}
\usepackage[figuresright]{rotating}

\title{An angular distribution analysis of  $\Lambda_b$ decays\footnote{Preprint numbers: 
ECT*-04-24, PCCF RI 0418}}

\author{O. Leitner\address{${\rm{ECT}}^{*}$, Strada delle Tabarelle, 286,
38050 Villazzano (Trento), Italy}%
        \thanks{leitner@ect.it},
        Z.J. Ajaltouni\address[MCSD]{Laboratoire de Physique Corpusculaire de Clermont-Ferrand, \\
IN2P3/CNRS, Universit\'e Blaise Pascal, 
F-63177 Aubi\`ere Cedex France}\thanks{ajaltouni@clermont.in2p3.fr}
        and 
        E. Conte\addressmark[MCSD]\thanks{conte@clermont.in2p3.fr}}
       
\begin{document}

\maketitle

\begin{abstract}
A  study of the angular distributions of the processes $\Lambda_b \to {\Lambda} V(1^-)$ with
$V ({J/\Psi,\rho}^0)$ is performed. Emphasis is put on the initial $\Lambda_b$ polarization
 and  the polarization density-matrices  are derived to perform tests of both Time-Reversal (TR) 
and $CP$ violation.
\end{abstract}
%
\section{$\boldsymbol{\Lambda_b}$ DECAY AND INTERMEDIATE RESONANCE ANALYSIS}
%
In the collisions, $p p \rightarrow  {\Lambda_b} + X$, the $\Lambda_b$ is 
produced with a  transverse polarization in a similar way than the 
ordinary hyperons. Let us define, 
${\vec N}_P$, the vector normal to the production plane by 
${\vec N}_P = (\vec{p}_1 \times \vec{p}_b)/ (|{\vec{p}_1 \times \vec{p}_b}|),$ 
where $\vec{p}_1$ and $\vec{p}_b$ are  the vector-momenta of one incident 
proton beam and $\Lambda_b$, respectively.
Let $({\Lambda_b}xyz)$ be the rest frame  of the $\Lambda_b$ 
particle. The quantization axis $({\Lambda_b}z)$ is chosen to be parallel 
to ${\vec N}_P$. The other orthogonal axis $({\Lambda_b}x)$ and $({\Lambda_b}y)$ 
are chosen arbitrarily in the production plane. In our analysis, 
the $({\Lambda_b}x)$ axis is taken parallel to the momentum $\vec {p}_1$. 
The spin projection, $M_i$, of the $\Lambda_b$ along the  transverse axis 
$({\Lambda_b}z)$ takes  the values   $\pm {1/2}$. The  polarization density 
matrix (PDM)~\cite{Leader:2001}, ${\rho}^{\Lambda_b}$, of the $\Lambda_b$ is a $(2 \times 2)$ 
hermitian matrix. Its elements, ${\rho}_{ii}^{\Lambda_b}$,  are real and 
$\sum_{i=1}^{2} {\rho}_{ii}^{\Lambda_b}= 1$. The  probability of having $\Lambda_b$  
produced with $M_i = \pm{1/2}$ is given by ${\rho}_{11}^{\Lambda_b}$ and 
${\rho}_{22}^{\Lambda_b}$, respectively. Finally, the initial $\Lambda_b$ polarization 
is  given by $\langle \vec{S}_{\Lambda_b} \cdot \vec{N}_P \rangle = {\cal P}_{\Lambda_b}
={\rho}_{11}^{\Lambda_b} - {\rho}_{22}^{\Lambda_b}.$

The decay amplitude, $A_0{(M_i)}$, for ${\Lambda_b}{(M_i)}  \to  
{\Lambda}{(\lambda_1)} V{(\lambda_2)}$
is obtained by applying the Wigner-Eckart theorem to the $S$-matrix
element in the framework of the Jacob-Wick helicity formalism~\cite{Jackson:1965}:
\begin{eqnarray}
 A_0{(M_i)} = \langle 1/2,M_i|S^{(0)}|p,\theta,\phi;\lambda_1,\lambda_2 \rangle 
= \mathcal{M}_{\Lambda_b}{(\lambda_1,\lambda_2)} D_{M_i M_f}^{{1/2} \star}{(\phi,\theta,0)}\ , \nonumber
\end{eqnarray}
where $\vec p = (p,\theta,\phi)$ is the vector-momentum of the hyperon 
$\Lambda$ in the $\Lambda_b$ frame. $\lambda_1$ and $\lambda_2$ are the 
respective helicities of $\Lambda$  and $V$ with the possible values $\lambda_1=\pm{1/2}$ 
and  $\lambda_2 =-1,0,+1$. The momentum projection along the 
$({\Delta})$ axis (parallel to $\vec{p}$) is given by $M_f =\lambda_1-\lambda_2=\pm{1/2}$.
The $M_f$ values  constrain those of $\lambda_1$ and $\lambda_2$ since, among six 
combinations, only  four are physical. 
The hadronic matrix element, $\mathcal{M}_{\Lambda_b}{(\lambda_1,\lambda_2)}$, 
contains  all the  decay dynamics. Finally, the Wigner matrix element,
$D_{M_i M_f}^j{(\phi,\theta,0)} = d_{M_i M_f}^j{(\theta)} {\exp{(-i{M_i}{\phi})}},$
 is expressed according to the Jackson convention~\cite{Jackson:1965}. 
By performing appropriate rotations and Lorentz boosts, we can study the decay 
of each resonance in its own helicity frame  such that the 
quantization axis is parallel to the resonance momentum in the $\Lambda_b$ 
frame i.e. $\overrightarrow{O_1z_1} || \vec{p}_{\Lambda}$ and 
$\overrightarrow{O_2z_2} || \vec{p}_{V}=-\vec {p}_{\Lambda}$.  Let us focus on the decays 
${\Lambda}{(\lambda_1)} \to  P{(\lambda_3)} {\pi}^-{(\lambda_4)}$  and 
$V{(\lambda_2)} \to  {\ell}^-{(\lambda_5)} {\ell}^+{(\lambda_6)}$  
or $V{(\lambda_2)} \to h^-{(\lambda_5)} h^+{(\lambda_6)}$:
in the $\Lambda$ helicity frame, the projection of the total angular momentum, 
$m_i$, along the proton momentum, $\vec{p}_P$, is given by $m_1=\lambda_3-\lambda_4 =\pm{1/2}$. 
In the vector meson helicity frame, this projection is equal to $m_2=\lambda_5-\lambda_6= -1,0,+1$ 
if leptons and $m_2=0$ if  $\pi$. The decay amplitude, $A_i(\lambda_i)$, of 
each resonance can be written similarly as $A_0(M_i)$ and  we have,
\begin{eqnarray}
A_1(\lambda_1) =\langle \lambda_1, m_1|S^{(1)}|p_1,{\theta}_1,{\phi}_1;\lambda_3,\lambda_4 \rangle =  
 \mathcal{M}_{\Lambda}{(\lambda_3,\lambda_4)} D^{{1/2} \star}_{\lambda_1 m_1}{(\phi_1,\theta_1,0)}\ , \nonumber \\
A_2(\lambda_2) =\langle  \lambda_2, m_2|S^{(2)}|p_2,{\theta}_2,{\phi}_2;\lambda_5,\lambda_6 \rangle  =  
 \mathcal{M}_{V}{(\lambda_5,\lambda_6)} D^{1 \star}_{\lambda_2 m_2}{(\phi_2,\theta_2,0)}\ , \nonumber
\end{eqnarray}
where  $\theta_1$ and  $\phi_1$ are respectively the polar and azimuthal angles 
of the proton momentum in the $\Lambda$ rest frame while $\theta_2$ and $\phi_2$ 
are those of ${\ell}^- {(h^-)}$ in the $V$ rest frame.
%
\subsection{Factorization procedure}
%
The effective interaction\footnote{All the terms of the 
effective interaction are extensively defined in literature.}, $\mathcal{H}^{eff}$, 
written as, $
\mathcal{H}^{eff}=\frac{G_{F}}{\sqrt{2}} V_{qb}V^{\star}_{qs} \sum_{i=1}^2 c_i(m_b) O_i(m_b)\ , 
$ 
gives the weak following amplitude factorized into,
\begin{eqnarray}
\mathcal{M}_{\Lambda_b}(\Lambda_b \to \Lambda V)= \frac{G_{F}}{\sqrt{2}} V_{qb}V^{\star}_{qs}
f_V E_V \Bigl( c_1 + \frac{c_2}{N_c^{eff}} \Bigr)
\langle \Lambda | \bar{s} \gamma_{\mu} (1-\gamma_{5}) b| \Lambda_b \rangle\ .\nonumber
\end{eqnarray} 
The CKM matrix elements, $V_{qb}V^{\star}_{qs}$, read as $V_{ub}V^{\star}_{us}$ 
and $V_{cb}V^{\star}_{cs}$, in case of $\Lambda_b \to \Lambda \rho$ and  
$\Lambda_b \to \Lambda J/\Psi$, respectively. The Wilson 
Coefficients, $c_i$, are equal to $c_1=-0.3$ and $c_2=+1.15$.
 Working in HQET, 
the final amplitude, 
$\mathcal{M}_{\Lambda_b}(\Lambda_b \to \Lambda V)$, depending on the helicity state, 
$(\lambda_{\Lambda},\lambda_V)$, reads as: $\mathcal{M}_{\Lambda_b} (\Lambda_b \to \Lambda V)=
\frac{G_{F}}{\sqrt{2}} V_{qb}V^{\star}_{qs} f_V E_V \Bigl( c_1 + \frac{c_2}{N_c^{eff}} \Bigr) \times$ 
\begin{align}
&-\frac{P_V}{E_V} \Biggl( \frac{m_{\Lambda_b}+m_{\Lambda}}
{E_{\Lambda}+m_{\Lambda}} F^-(q^2) + 2 F_2(q^2) \Biggl)\ ; &
 \frac{1}{\sqrt{2}} \Biggl( \frac{P_{V}}
{E_{\Lambda}+m_{\Lambda}}F^-(q^2)+   F^+(q^2) \Biggl)\ ; \nonumber \\
&\frac{1}{\sqrt{2}}  \Biggl( \frac{P_{V}}
{E_{\Lambda}+m_{\Lambda}}F^-(q^2) - F^+(q^2) \Biggl)\ ; &
 \Biggl(F^+(q^2) + \frac{P^2_{V}}
{E_V (E_V+m_{\Lambda})} F^-(q^2) \Biggl)\ ;\nonumber
\end{align}
respectively, for $(\lambda_{\Lambda},\lambda_V)=(\frac12,0);(-\frac12,-1);(\frac12,1)$ and $(-\frac12,0)$.
The $q^2$ dependence of the transition form factors\footnote{We define $F^{\pm}(q^2)=F_1(q^2)\pm F_2(q^2)$, for 
convenience.}, $F_i(q^2)$, or $(F^{\pm}(q^2))$,  
resulting from QCDSR and HQET~\cite{Huang:1998ek} takes the form 
$
F_i(q^2)= F(0)/\bigl(1- a\frac{q^2}{m^2_{\Lambda_b}} + b\frac{q^4}{m^4_{\Lambda_b}}\bigr),
$
where  $(0.462,-0.0182,-1.76\!\times\!10^{-4})$ and $(-0.077,-0.0685,1.46\!\times\!10^{-3})$ 
correspond to $(F(0),a,b)$ in case of $F_1(q^2)$ and $F_2(q^2)$, respectively. We refer 
to the PDG~\cite{Eidelman:2004wy} for all the numerical values used in our analysis.
%
\subsection{Angular distributions}
%
First the  $\Lambda_b  \to  \Lambda V$ decay:
writing the hadronic matrix element, $\mathcal{M}_{\Lambda_b}(\lambda_1,\lambda_2)$,  
into two parameters according to the final helicity such as,
$
{|\mathcal{M}_{\Lambda_b} (\pm1/2)|}^2 = \   {|\mathcal{M}_{\Lambda_b}(\pm1/2,0)|}^2 
+ {|\mathcal{M}_{\Lambda_b}(\mp1/2,\mp1)|}^2,
$
and by introducing the  helicity asymmetry parameter, ${\alpha}_{As}^{\Lambda_b}$, 
defined by, $
 \alpha_{AS}^{\Lambda_b} = ({|\mathcal{M}_{\Lambda_b}{(+1/2)}|}^2 -
 {|\mathcal{M}_{\Lambda_b}{(-1/2)}|}^2)/({|\mathcal{M}_{\Lambda_b}{(+1/2)}|}^2 +
 {|\mathcal{M}_{\Lambda_b}{(-1/2)}|}^2),$
the final angular distribution, $W{(\theta, \phi)}$, which can be expressed as, 
\begin{eqnarray}
W{(\theta, \phi)}  \propto    1 + {\alpha_{AS}^{\Lambda_b}}{\cal P}_{\Lambda_b}{\cos\theta} +
 2{\alpha_{AS}^{\Lambda_b}} \Re e \Big[ {\rho}_{ij}^{\Lambda_b}{\exp{(-i\phi)}} \Big]\sin{\theta}\ , \nonumber
\end{eqnarray} 
puts into evidence the parity violation. Regarding the $\Lambda \to  P {\pi}^-$ decay,  the 
general formula for proton angular distributions, 
$W_1{(\theta_1,\phi_1)}$, in the $\Lambda$ rest-frame reads as,
\begin{multline}
W_1{(\theta_1,\phi_1)} \propto \\
\frac12 \Biggl\{ ({\rho}_{ii}^{\Lambda}+{\rho}_{jj}^{\Lambda})+
({\rho}_{ii}^{\Lambda}-{\rho}_{jj}^{\Lambda}) 
\alpha_{AS}^{\Lambda} \cos\theta_1   
 -  \frac{\pi}{2}{\cal P}_{\Lambda_b}\alpha_{AS}^{\Lambda}
  \Re e \Big[{\rho}_{ij}^{\Lambda} \exp{(i\phi_1)} \Big] \sin{\theta_1}\Biggr\} \ ,\nonumber
\end{multline} 
whereas for the $V \to  {\ell}^+ {\ell}^-  (h^+ h^-)$ decay,  the angular distributions, $W_2{(\theta_2,\phi_2)}$, 
in the $V$ rest-frame, is given by,
\begin{multline}
W_2{(\theta_2, \phi_2)}    \propto     
  ({\rho}_{ii}^{V}+{\rho}_{jj}^{V})(G^V_{00}(\theta_2,\phi_2)+G^V_{\pm 1 \pm1}(\theta_2,\phi_2)) \nonumber
\\ -\frac{\pi}{4}{\cal P}_{\Lambda_b}  \Re e \Big[{\rho}_{ij}^V 
\exp{(i\phi_2)} \Big] \sin{2\theta_2}\ ,\nonumber
\end{multline}
where  the PDM elements, ${\rho}_{ij}^{V(\Lambda)}$, of $V(\Lambda)$ and $G^V_{\lambda_2,\lambda_2^{\prime}}
(\theta_2,\phi_2)$ are given in Ref.~\cite{Ajaltouni:2004zj}.
%
\section{RESULTS}
%
 The helicity asymmetry parameter, $\alpha_{As}^{\Lambda_b}$, 
 takes the  values 
${\alpha}^{\Lambda_b}_{AS}= 98.8\%$  for  $\Lambda_b \to \Lambda \rho^0$ and
${\alpha}^{\Lambda_b}_{AS}= 77.7\%  \;\;{\rm for}\;\; \Lambda_b \to \Lambda J/\Psi.$
The $\Lambda$-polarization, $\mathcal{P}_{\Lambda}=\rho_{11}^{\Lambda}-\rho_{22}^{\Lambda}$,
 can be computed in both decays. After normalization of $\mathcal{P}_{\Lambda}$,  we obtain the values, 
$\mathcal{P}_{\Lambda}=+31\%$, and $\mathcal{P}_{\Lambda}=-9\%$, for the $\Lambda {\rho}^0$ 
and $\Lambda J/{\psi}$ channels, respectively.
Another  parameter concerning the spin state of the intermediate resonances is
the density matrix element, $\rho^V_{ij}$. Let us focus 
on $\rho^V_{00}$ which is  related to the longitudinal polarization of 
the vector meson V. After calculation, $65.5\%$ and $55.5\%$
are the results for $\rho^V_{00}$ in case of  $\Lambda_b \to 
\Lambda {\rho}^0$ and $\Lambda_b \to \Lambda J/{\psi}$, respectively.
It is important to notice that these parameters, ${\alpha}^{\Lambda_b}_{AS}$ and 
$\rho^V_{ij},$ (as well as $\rho^{\Lambda}_{ij}$) govern entirely the angular distributions, 
$W_i(\theta_i,\phi_i)$, of the final particles. 
 Performing all the calculations and keeping the number of color,
$N_c^{eff}$, to vary between  2 and 3 as it is suggested by the factorization hypothesis, we
obtain the  branching ratio results $
\mathcal{BR}(\Lambda_b \to \Lambda \rho^0) =  (34.0\ , 11.4\ , 3.1) \times 10^{-8}$ and
$\mathcal{BR}(\Lambda_b \to \Lambda J/\psi) =  (12.5\ , 4.4\ ,1.2) \times 10^{-4},$
respectively for $N_c^{eff}=2, 2.5, 3$. 
For comparison, $\mathcal{BR}^{exp}(\Lambda_b \to \Lambda J/\psi)
=(4.7\pm2.1 \pm 1.9)\times 10^{-4}$~\cite{Eidelman:2004wy}.
%
\section{CONCLUSIONS}
%
Calculations of the angular distributions as well as branching ratios 
of the process $\Lambda_b \to \Lambda V$ with $\Lambda \to P {\pi}^-$ 
and  $V \to {\ell}^+ {\ell}^-(h^+ h^-)$  have been performed by 
using the helicity formalism.  The initial 
polarization, ${\cal P}_{\Lambda_b}$, appears  explicitly in the polar angle 
distribution of the $\Lambda$ hyperon in the $\Lambda_b$ rest-frame. 
Similarly,  the azimuthal angle distributions of both proton and ${\ell}^-$ 
in the $\Lambda$ and $V$ frames, respectively, depend directly on the 
$\Lambda_b$ polarization. A first computation of the asymmetry 
parameter, ${\alpha}_{As}$, in $\Lambda_b$ decays into $\Lambda V(1^-)$ has 
been performed as well as the longitudinal polarization of the vector meson, 
${\rho}_{00}^V$, which is shown to be dominant $(\ge 56\%)$.
Looking for TR violation effects in baryon decays 
provides us  a complementary test of 
$CP$ violation by assuming the correctness of the $CPT$ theorem. 
In particular, triple product correlations, which are $T$-odd, 
can be investigated in $\Lambda_b$ decays.
\begin{figure}
\begin{center}
\includegraphics*[width=0.325\columnwidth]{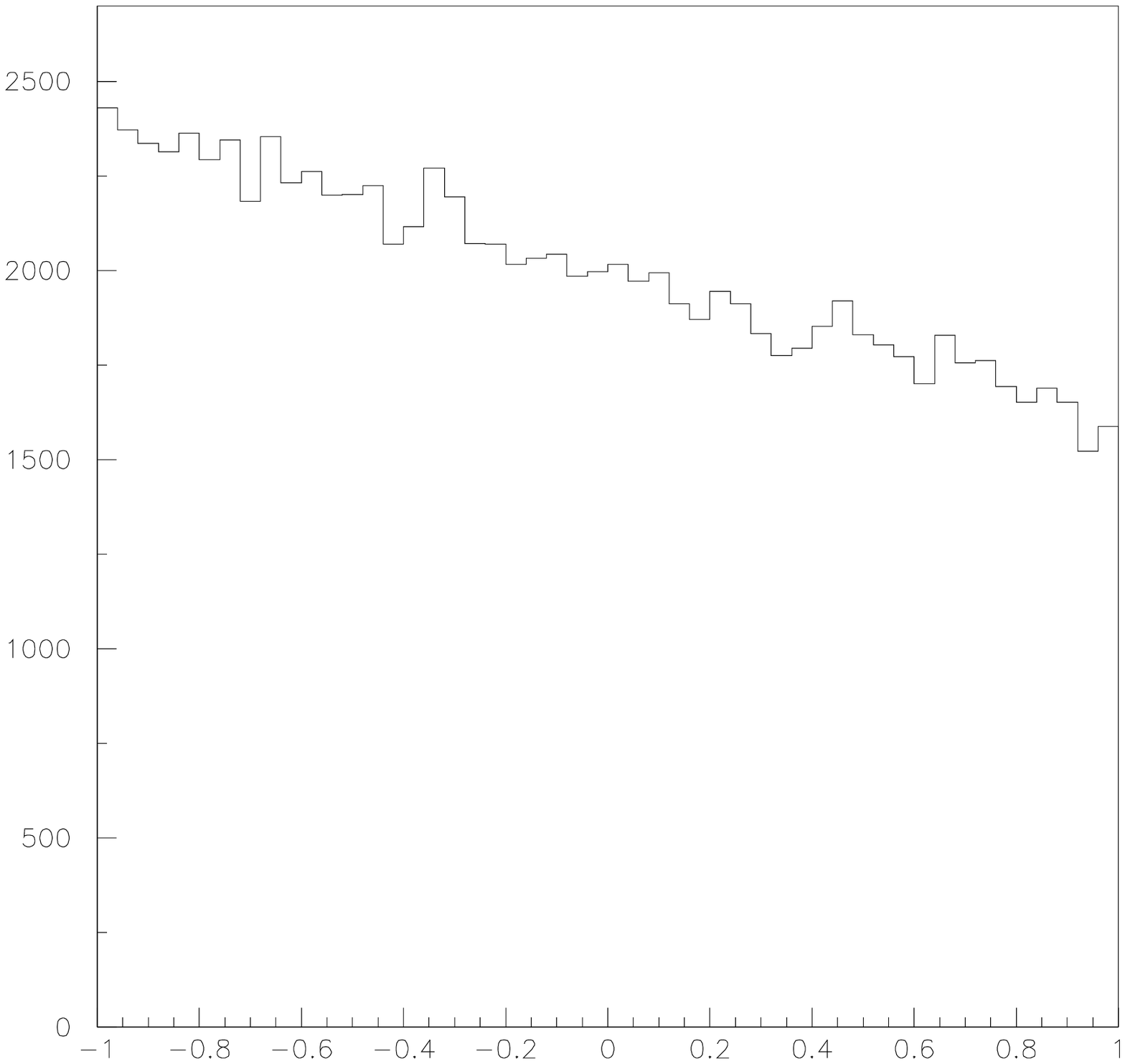}
\includegraphics*[width=0.325\columnwidth]{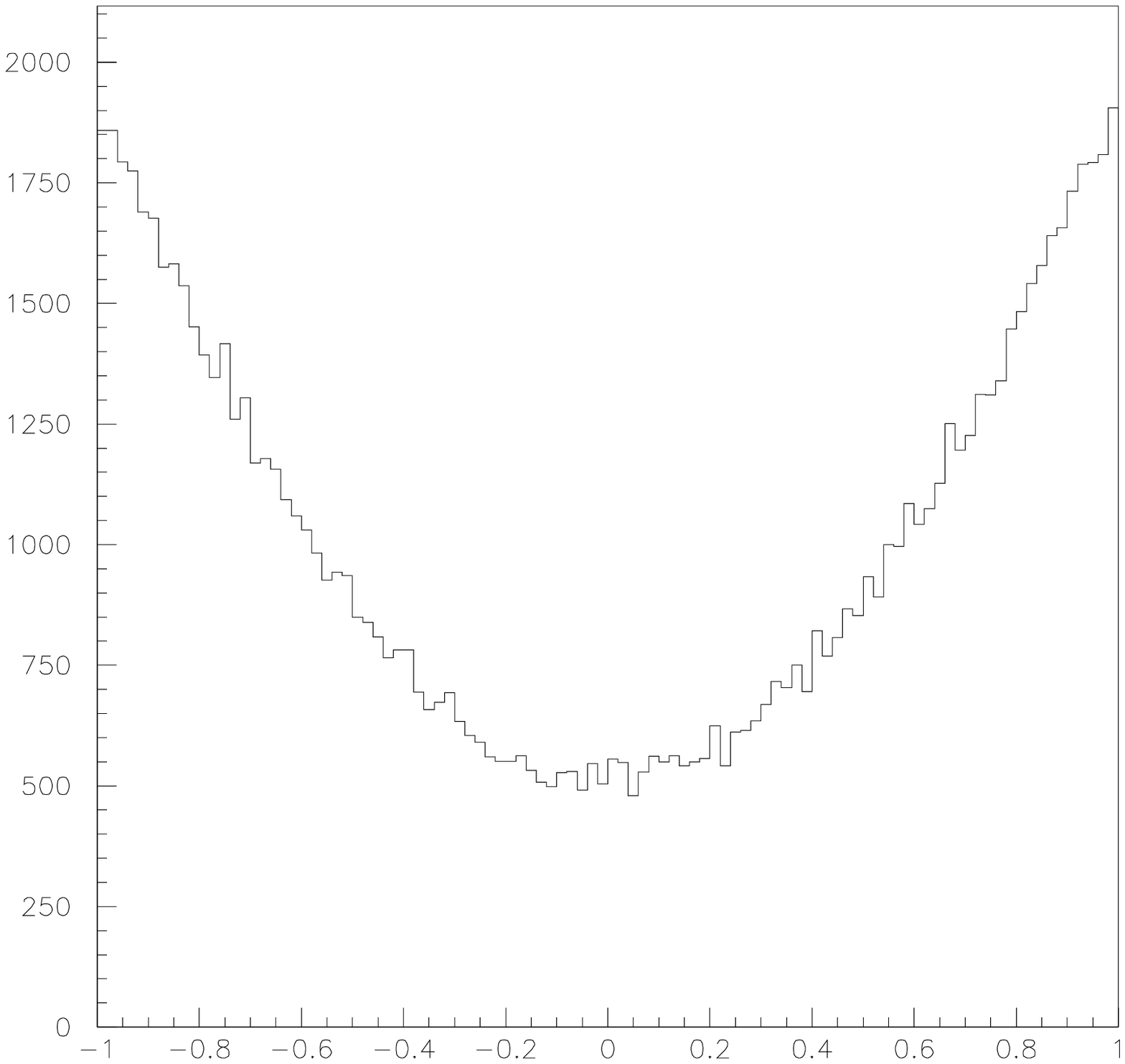}
\includegraphics*[width=0.325\columnwidth]{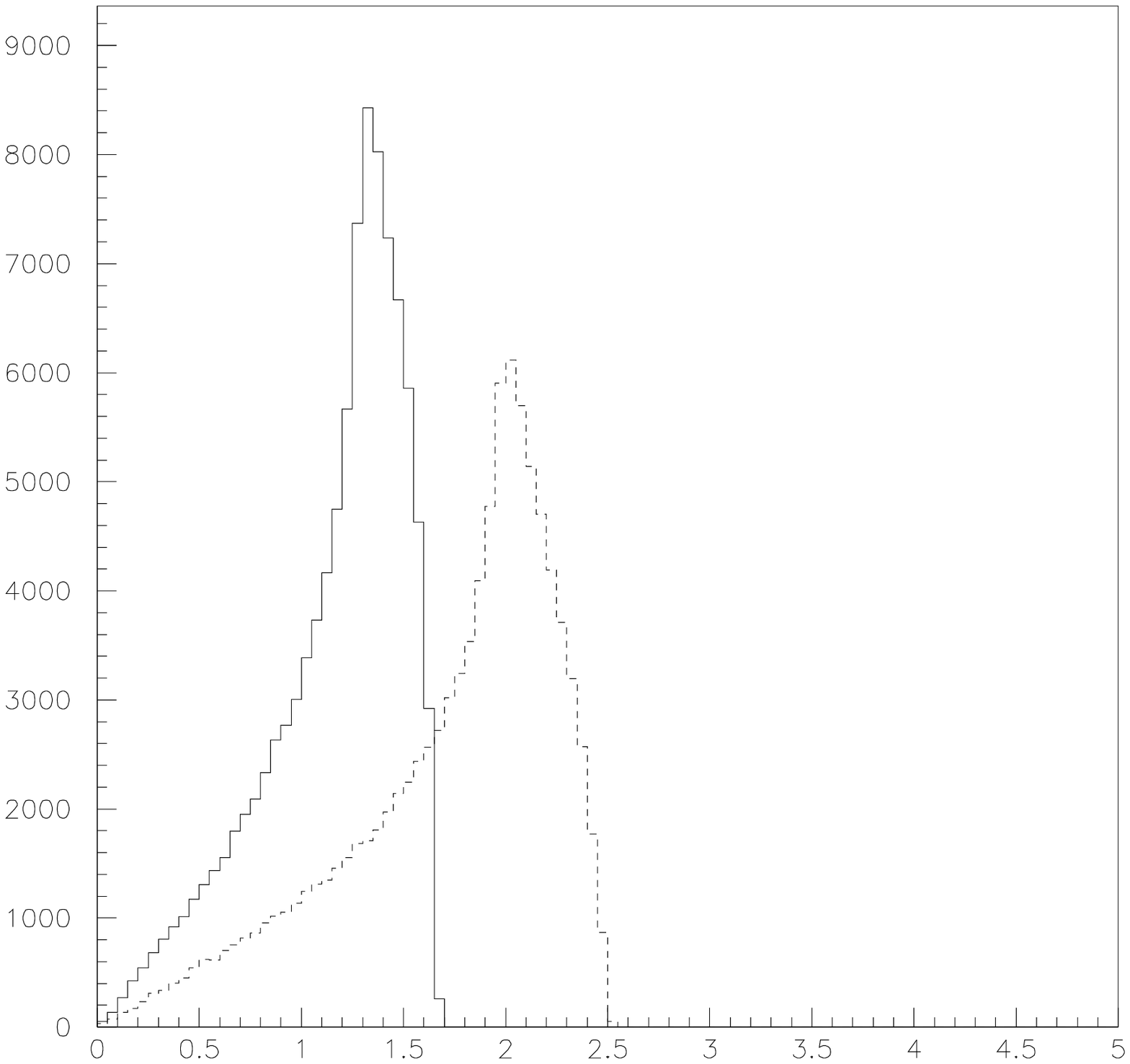}
\includegraphics*[width=0.325\columnwidth]{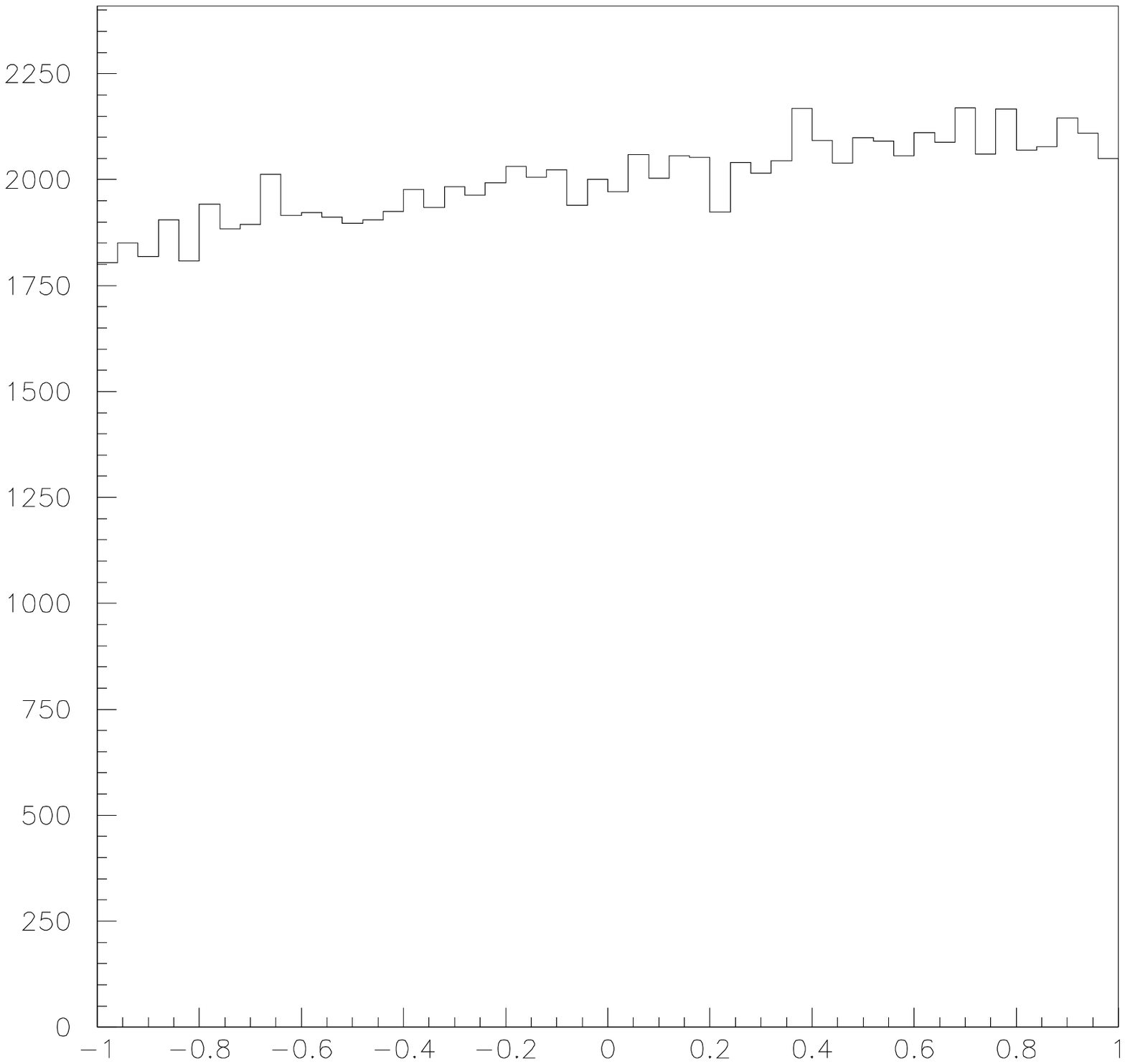}
\includegraphics*[width=0.325\columnwidth]{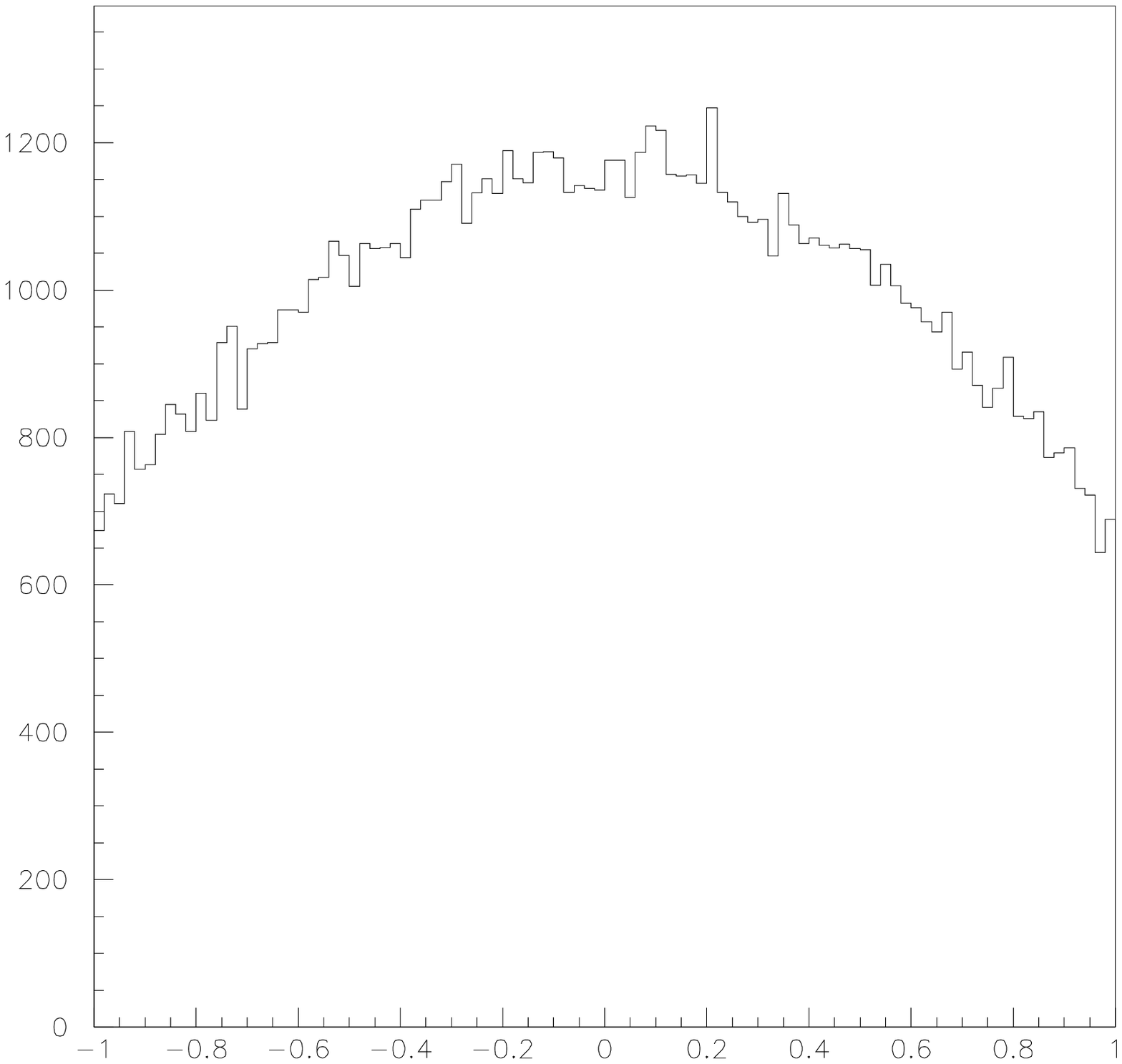}
\includegraphics*[width=0.325\columnwidth]{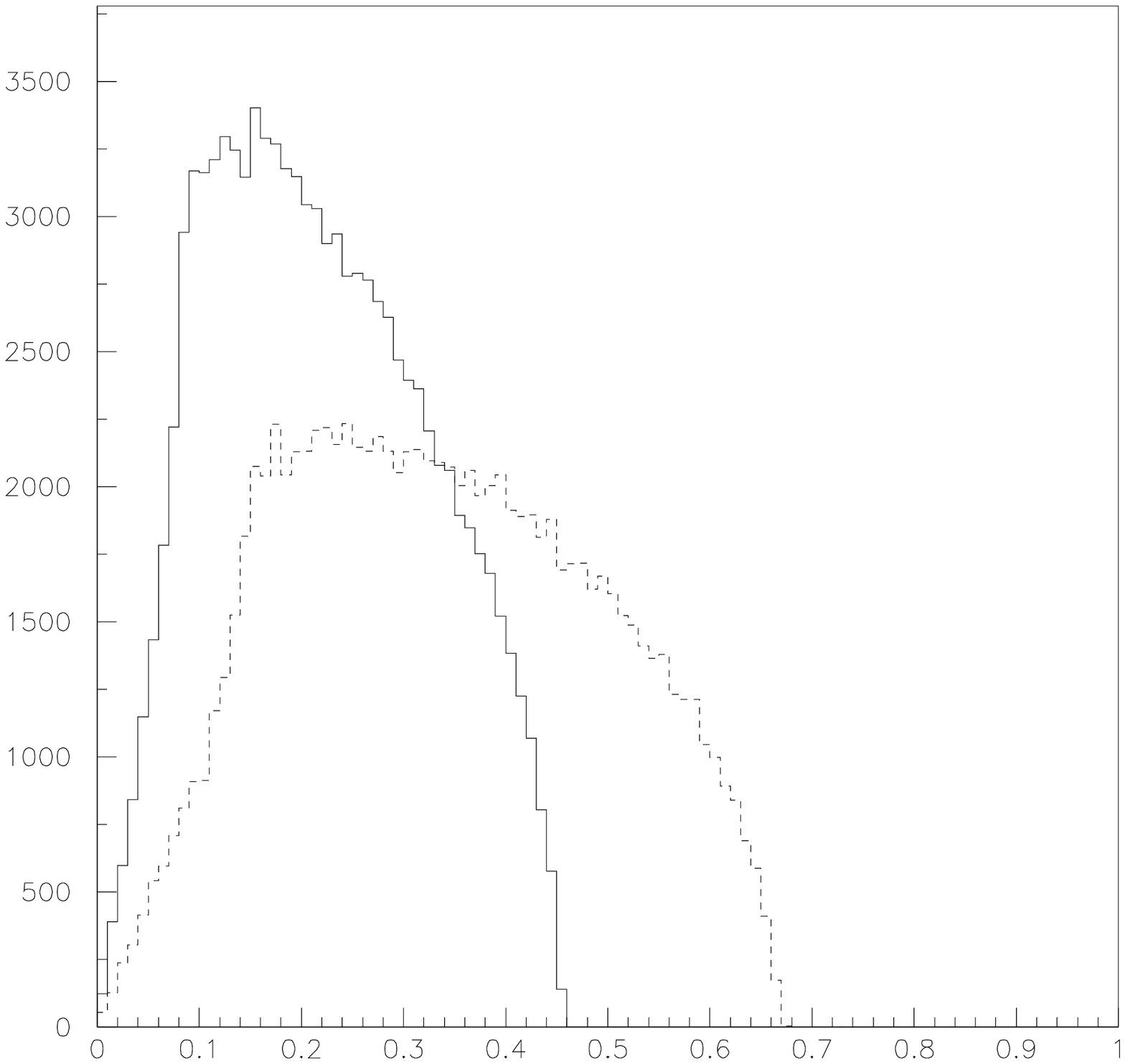}
\caption{First two columns: polar angular distributions in the intermediate resonance rest-frames 
for  the following values of $\mathcal{P}_{\Lambda}$ and ${\rho}_{00}^V$ parameters:
$(31\%, 65.5\%)$ in the case of $\Lambda_b \to \Lambda {\rho}^0$ (upper histograms: $\cos {\theta}_P$ 
(left side) and $\cos {\theta}_{{\pi}^-}$ (right side)), and $(-9\%, 55.5\%)$ in the case of
$\Lambda_b \to \Lambda J/{\psi}$ (lower histograms: $\cos {\theta}_P$ (left side) and 
$\cos {\theta}_{{\mu}^-}$ (right side)). \newline
Third column: proton and pion ($\Lambda$ daughters) transverse momentum, $P_{\perp}$, in the $\Lambda_b$ rest-frame, 
 in the case of $\Lambda {\rho}^0$ channel (dashed line) 
and $\Lambda J/{\psi}$ channel (full line), respectively. 
Upper histogram for  proton $P_{\perp}$-spectra and lower histogram for pion $P_{\perp}$-spectra.}
\end{center}
\end{figure}

\end{document}